\documentclass[amsmath,amssymb,twocolumn,superscriptaddress,preprintnumbers,nofootinbib,prd]{revtex4-1}

\usepackage[utf8]{inputenc}
\usepackage{amssymb}
\usepackage{cancel}
\usepackage{amsmath}
\usepackage{verbatim}
\usepackage{braket}
\usepackage{graphicx}
\usepackage{siunitx} 

\usepackage{hyperref}
\usepackage{cleveref} %This package has to go last

\DeclareMathOperator{\Br}{Br}
\DeclareMathOperator{\erf}{erf}
\newcommand{\nnnl}{\nonumber\\}

\graphicspath{{figures}}

\begin{document}
%%%%%%%%%%%%%%%%%%%%%%

\vspace*{-0.7cm}

\title{Sub-GeV dark matter from cosmic ray bremsstrahlung in the atmosphere}

\author{Branden Aitken}
 \email{brandenaitken@uvic.ca}
\affiliation{Department of Physics and Astronomy, University of Victoria, Victoria BC V8P 5C2, Canada}

\author{Peter Reimitz}
 \email{peter@if.usp.br}
\affiliation{Department of Physics and Astronomy, University of Victoria, Victoria BC V8P 5C2, Canada}
\affiliation{Instituto de F\'{i}sica,
Universidade de S\~{a}o Paulo, 05508-090 S\~{a}o Paulo, SP, Brasil}

\author{Adam Ritz}
 \email{aritz@uvic.ca}
\affiliation{Department of Physics and Astronomy, University of Victoria, Victoria BC V8P 5C2, Canada}

\begin{abstract}
We explore the sensitivity of neutrino observatories and direct dark matter detection experiments to boosted sub-GeV dark matter produced by inelastic cosmic ray collisions in the atmosphere. We revisit earlier approaches and extend the sensitivity to higher mass by modeling the proton bremsstrahlung production mode via initial state radiation. For vector-mediated dark matter models, the peak of the cosmic ray flux allows for enhanced DM production for mediator masses near the $\rho/\omega$ resonances. We determine and compare the ensuing sensitivity of direct detection experiments LZ and PandaX-4T and the neutrino detectors Borexino and Super-K.  
\end{abstract}

\maketitle

%%%%%%%%%%%%%%%%%%%%%%%%%%%%%%%%%%%%%%%%%%%%%%%%%%%%%%
\section{Introduction}
\label{sec:intro}
%%%%%%%%%%%%%%%%%%%%%%%%%%%%%%%%%%%%%%%%%%%%%%%%%%%%%%

Models of dark matter (DM) that utilize thermal freezeout in the early universe remain a compelling benchmark for experimental searches because their relic abundance is insensitive to early universe initial conditions. Extensive efforts have been made in recent years to explore the full mass range for thermal relic DM in the galactic halo, including lighter sub-GeV candidates that may elude conventional direct detection experiments due to nuclear recoil energy thresholds optimized for weak-scale DM. In addition to progress with low threshold direct detection experiments (see e.g. \cite{Essig:2022dfa}), this effort has driven a large and complementary program of accelerator-based probes of sub-GeV DM models, which can exploit the couplings inherent in thermal relic freezeout to probe DM as part of an extended dark sector in a cosmologically relevant kinematic regime (see e.g. \cite{Beacham:2019nyx,Gori:2022vri,Krnjaic:2022ozp}).

A related strategy to explore sub-GeV DM uses the analogy with the atmospheric production of neutrinos from inelastic cosmic ray scattering \cite{Alvey:2019zaa}, as this leads to a much harder kinematic spectrum for sub-GeV DM and thus more energetic recoil signals in detectors. This `atmospheric beam dump' also has the novel feature of scanning over a wide range of `beam energies' due to the extended nature of the cosmic ray spectrum. Initial work focused on the production of DM and millicharge particles through meson decay, which is the dominant low mass production mode, and subsequent extensions considered additional production channels \cite{Alvey:2019zaa,Plestid:2020kdm,Wu:2024iqm,Arguelles:2019ziu, Du:2022hms, Du:2023hsv} including bremsstrahlung. (See also \cite{An:2017ojc,An:2021qdl,Emken:2021lgc,Emken:2024nox,Bringmann:2018cvk,Ema:2018bih,Cappiello:2019qsw,Acevedo:2026xol} for alternative mechanisms for producing boosted DM.)

In this paper, we revisit the atmospheric production mechanism for DM and extend the reach in mass for dark photon mediated benchmark models of sub-GeV DM. Starting from the proton component of the cosmic ray spectrum, we utilize a recently developed formalism for proton bremsstrahlung that focuses on initial state radiation. This formalism naturally incorporates precision fits for the proton monopole and dipole form factors that allow for a significant enhancement in DM production via mixing with vector resonances.

The remainder of this paper is organized as follows. In \Cref{sec:crdm}, we review the atmospheric beam dump for cosmic rays and the resulting DM kinematics, and in \Cref{sec:pbrem} we provide more details of the bremsstrahlung production cross section. \Cref{sec:sens} develops the modeling of scattering signatures in DM direct detection and large volume neutrino experiments, and the resulting sensitivity. Concluding remarks are provided in \Cref{sec:disc}.

%%%%%%%%%%%%%%%%%%%%%%%%%%%%%%%%%%%%%%%%%%%%%%%%%%%%%%
\section{Atmospheric DM via cosmic ray scattering}
\label{sec:crdm}
%%%%%%%%%%%%%%%%%%%%%%%%%%%%%%%%%%%%%%%%%%%%%%%%%%%%%%

The primary advantage of atmospheric production via cosmic ray interactions is that it creates DM  particles that are highly boosted relative to characteristic speeds in the galactic halo.    
This enables lighter DM particles (in the sub-GeV range) to create nuclear or electron recoil signatures of similar strength to their more traditional WIMP counterparts and allows the use of direct detection and neutrino experiments whose energy detection thresholds have historically been unfavorable for the detection of sub-GeV DM.

A key component of this production model is the flux $\Phi_p$ of cosmic ray protons. Additional cosmic ray species have considerably lower fluxes \cite{Boschini:2017fxq} and will not be included here. Since we will focus on the bremsstrahlung production mode in a regime with relativistic kinematics, we can limit our attention to the energy spectrum above $E_p \gtrsim \mathcal{O}(10)$~GeV, which is well described by a simple power law $\frac{d \Phi_p}{dE_p} \propto E_p^{-2.7}$ \cite{ParticleDataGroup:2024cfk},
\begin{align}
    \frac{d^2 \Phi_p}{dE_p d\Omega_p} \approx \frac{1.3}{\si{cm^2 . s . sr . GeV}}\left(\frac{E_p}{\si{GeV}}\right)^{-2.7}.
\end{align}
  Concretely, to ensure relativistic kinematics for proton bremsstrahlung on nuclei in the atmosphere, we impose a low energy cutoff on this flux when 
 \begin{align}\label{eqn:gammacut}
  \gamma_\text{cm}^2 \gtrsim  5, 
  \end{align}
  corresponding to $E_p \gtrsim 9 m_p$ in the lab frame. %

\begin{figure*}[t]
    \centering
    \includegraphics{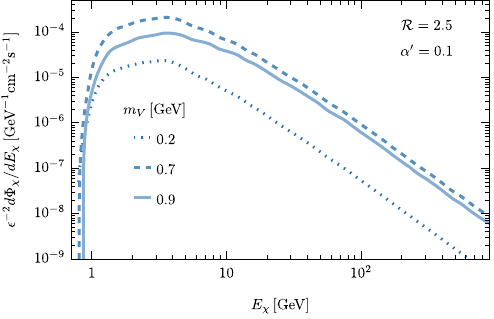}
    \includegraphics[scale=1.01]{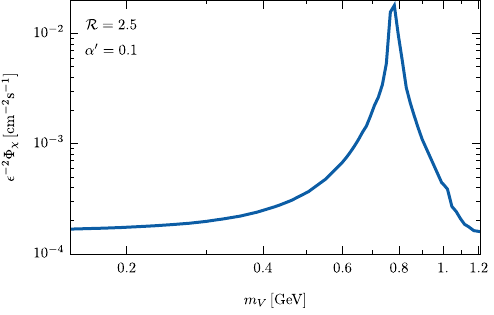}
    \caption{The flux distribution of cosmic ray DM computed from \cref{eqn:dflux}, which scales as $\epsilon^2$. (Left) The differential flux is shown with mass points chosen to emphasize the impact of the $\rho/\omega$ resonance. The steep falloff below $E_\chi \sim 1 \, \si{GeV}$ is a direct result of the cutoff applied to the cosmic ray flux in \cref{eqn:gammacut}. (Right) The total flux incident on the detector for a given vector mass. The $\rho/\omega$ resonant peak is a result of the form factors used in \cite{Kling:2025udr}, while the small inflection at the $\phi$ mass reflects the more conservative treatment of the impact of mixing with this resonance.}
    \label{fig:diff}
\end{figure*}

Given a cosmic ray proton of fixed energy, we can compute the rate of dark bremsstrahlung processes using the formalism for proton initial state radiation (ISR) developed in \cite{Foroughi-Abari:2021zbm, Foroughi-Abari:2024xlj, Kling:2025udr}, which is summarized in more detail in \cref{sec:pbrem}. For models of DM  that utilize a dark vector mediator, this results in a dark vector energy spectrum, with the vector bosons undergoing a two-body decay to DM. In the lab frame, this production rate can then be convolved with the original cosmic ray spectrum \cite{Wu:2024iqm, Arguelles:2019ziu, Du:2022hms, Du:2023hsv}, to yield the differential DM flux. Since the mean free path of a cosmic ray proton in the atmosphere is much shorter than the height of the atmosphere,
we can assume that all protons scatter, and thus the DM flux follows directly from the splitting probability to produce the dark vector,
\begin{align}\label{eqn:dflux}
    \frac{d^2 \Phi_\chi}{d E_\chi d\Omega_p} = \,  2 &\int_{9 m_p}^{\infty} d E_p \frac{d^2 \Phi_p}{d E_p d\Omega_p} \nnnl
    \times &\int_{E_V^-}^{E_V^+} d E_V 
    \frac{1}{\sigma_{pp}^\text{tot}} 
    \frac{d \sigma_\text{br}}{d E_V} \frac{\Br(V \to \chi \overline{\chi})}{E_\chi^+ - E_\chi^-}.
\end{align}
In this expression, considering both $\chi$ and $\overline{\chi}$ gives the factor of 2 in front of the integral, and the differential cosmic ray proton flux is then integrated from our imposed relativistic lower cutoff of $E_p = 9m_p$ to infinity. This lower cutoff allows us to ignore nuclear effects in scattering. As described in \cref{sec:pbrem}, the production rate is determined by the differential ISR cross section $d \sigma_\text{br}/d E_V$ for $p+p \rightarrow V(\rightarrow \chi\bar{\chi}) + X$, with $X$ any inelastic final state, and we obtain the probability to radiate a vector by dividing by the underlying total cross section $\sigma_{pp}^{\rm tot}$ for proton-proton scattering.

Attenuation effects in the atmosphere or in the Earth that arise once the DM particle is produced can be accounted for by restricting the $d\Omega$ integral to be over the appropriate angular range that reaches the detector.  These effects are relevant when the DM mean free path is comparable to or less than the radius of the Earth. However, the range of cross-section constraints on DM produced via bremsstrahlung will allow us to safely neglect this source of attenuation \cite{Wu:2024iqm}.

Finally, the quantities $E_\chi^\pm$ and $E_V^\pm$ in \eqref{eqn:dflux} account for the boost of the $V$ decay to the Earth's rest frame \cite{Wu:2024iqm}, and lead to the box distribution of height $1/(E_\chi^+ - E_\chi^-)$. We can write $E_\chi^\pm$ in terms of the vector rest frame variables (denoted here with primes) as 
\begin{align}
    E_\chi^\pm = \gamma\left(E_\chi' \pm \beta \vert p'_\chi \vert \right).
\end{align}
For a two-body decay, we have $m_V = 2 E_\chi'$ and ${p_\chi'}^2 = m_V^2/4 - m_\chi^2$, allowing us to rewrite the vector integration bounds as a function of $E_\chi$ for \cref{eqn:dflux}.  This relation can be found by solving $E_\chi = E_\chi^\pm$,
\begin{align}
    E_V^\pm = E_\chi \mathcal{R}^2 \pm \mathcal{R} \sqrt{(E_\chi^2 - m_\chi^2)(\mathcal{R}^2 - 4)},
\end{align}
where $\mathcal{R} \equiv m_V / m_\chi$.

%%%%%%%%%%%%%%%%%%%%%%%%%%%%%%%%%%%%%%%%%%%%%%%%%%%%%%
\section{DM production via proton bremsstrahlung}
\label{sec:pbrem}
%%%%%%%%%%%%%%%%%%%%%%%%%%%%%%%%%%%%%%%%%%%%%%%%%%%%%%

We will focus on DM models with a kinetically mixed dark photon $A_\mu'$ mediating the interaction with Standard Model degrees of freedom. The current interaction takes the generic form,
\begin{align}
 {\cal L} &\supset \epsilon e \sum_f q_f \bar{f} \gamma^\mu A'_\mu f,
\end{align}
with $\epsilon$ parameterizing the overall mixing strength relative to the EM current and $q_f$ the relevant fermion charge.

Production of dark vectors in the collision of cosmic ray protons in the atmosphere has been outlined above, and depends on the differential bremsstrahlung cross section. 
The determination of this cross section closely follows the framework developed in \cite{Foroughi-Abari:2021zbm,Foroughi-Abari:2024xlj,Kling:2025udr}, and a conceptually similar implementation in \cite{Gorbunov:2024vrc}. In \cite{Foroughi-Abari:2021zbm}, the quasi-real approximation was introduced, which focused on a systematic approximation to initial state radiation in place of variants of the Weiz\"{a}cker-Williams approximation, as in \cite{Blumlein:2013cua}. The framework in \cite{Foroughi-Abari:2021zbm} was subsequently improved in \cite{Foroughi-Abari:2024xlj} by incorporating the Dawson prescription to remove divergences at low dark vector masses and by extending the treatment of hadronic structure through the inclusion of a dipole form factor in addition to the monopole form factor. Most recently, \cite{Kling:2025udr} introduced an extended set of form-factor parameterizations designed to provide increased flexibility for different BSM scenarios, while also enabling a consistent treatment of both proton and neutron bremsstrahlung, together with a systematic uncertainty estimate.

In the present work, we build directly on the most recent developments of \cite{Kling:2025udr}, for which the differential cross section can be factorized as follows,
\begin{align}
    \frac{d^2\sigma_{br}}{d E_V d\theta} \approx 2\epsilon^2 p_T z \sigma_{\rm NSD}(s') w(z,p_T^2) |F_p(m_V^2)|^2 K(p_V),
\end{align}
where to simplify the presentation we have assumed relativistic and collinear kinematics, so that the lab frame dark vector 3-momentum $p_V \approx E_V$, the transverse momentum $p_T \approx p_V \theta$, where $\theta$ is the angle of $V$ relative to the incoming proton, and $z\approx E_V/E_p$ is the fraction of energy carried by the vector.  The angular integral over $\theta$ from $0$ and $\pi/2$ is performed without assuming collinear kinematics prior to the computation of \cref{eqn:dflux}.

As detailed in \cite{Foroughi-Abari:2021zbm,Foroughi-Abari:2024xlj,Kling:2025udr} the differential cross section describes ISR parametrized by a splitting function $w(z,p_T^2)$ and an underlying non-single diffractive $pp$ cross section $\sigma_{\rm NSD}$ that is evaluated at a reduced center of mass energy $s' \approx (1-z)s$. We use  the parameterization of $\sigma_{\rm NSD}$ from \cite{Likhoded:2010pc}. The splitting functions $w(z,p_T^2)$ and associated proton form factors $F_p(m_V^2)$ are described in detail in \cite{Foroughi-Abari:2024xlj,Kling:2025udr}. Finally, the factor $K(p_V)$ implements kinematic constraints for the validity of this ISR approximation. In practice, this requires that the proton scattering is relativistic, which is implemented above with the lower cutoff in the cosmic ray spectrum, and approximate collinearity of the vector. 

A viable sub-GeV DM model can be introduced via the addition of scalar or fermionic fields $\chi$ charged under the vector mediator $A'_\mu$,
\begin{align}
 {\cal L} &\supset g' J_D^\mu A'_\mu,
\end{align}
 where the dark current is $J_D^\mu = i\chi^\dagger{\stackrel{\leftrightarrow}{\partial_\mu}}\chi$ for scalar DM, which we take as our default case below. For reference, $J_D^\mu = i\bar{\chi}_1 \gamma^\mu \chi_2$ for pseudo-degenerate fermions, with a small mass splitting between the DM candidate $\chi_1$ and $\chi_2$ required to avoid constraints on s-wave annihilation. In all cases, we assume that $g'$ (and the associated dark fine structure constant $\alpha' \equiv (g')^2/(4\pi)$) is sufficiently large that Br($V\rightarrow \chi\overline{\chi})\approx 1$ when kinematically allowed.

\begin{figure}[t]
    \centering
    \includegraphics{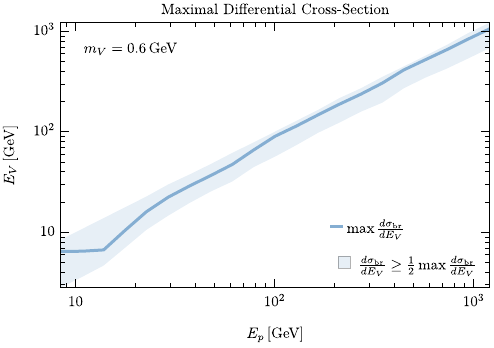}
    \caption{The peak bremsstrahlung production rate occurs for $E_V$ close to $E_p$. The solid contour in the figure tracks the peak, with the band showing the narrow range over which the differential cross section is no less than half the peak value. }
    \label{fig:bremxsec}
\end{figure}

With this formalism in hand, we adopt the same configuration choices as in \cite{Kling:2025udr} and generate differential cross sections for a range of dark vector masses, $pp$ center-of-mass energies, and dark vector momenta. The resulting lab frame DM flux is exhibited in \cref{fig:diff} and we will present all results in the lab frame for consistency below.  As expected, the vector energy is strongly correlated with the energy of the incoming proton.  We can see in \cref{fig:bremxsec} that the peak production occurs when $E_V \sim E_p$ with a steep fall-off for $E_V$ beyond the peak.  The low energy tail is wider producing an asymmetric distribution.

%%%%%%%%%%%%%%%%%%%%%%%%%%%%%%%%%%%%%%%%%%%%%%%%%%%%%%
\section{Sensitivity of DM and Neutrino Detectors}
\label{sec:sens}

\begin{figure*}[t]
    \centering
     \includegraphics{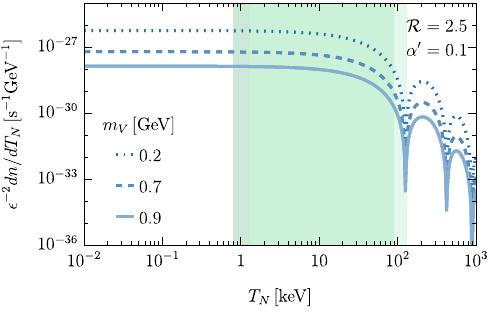}
    \includegraphics[scale=1.02]{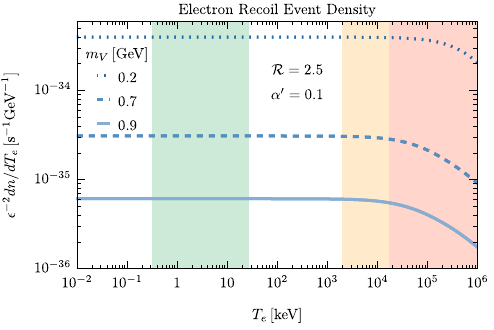}
    \caption{The recoil event density \cref{eqn:eventdensity} evaluated with the differential cross section of \cref{eqn:diffxsec} with the appropriate form factor substitution for nuclear scattering (left) and electron scattering (right).  The bands show the detection thresholds of each detector. The relevant nuclear recoil signatures are for LZ (light green) and PandaX-4T (dark green) with their regions having significant overlap.  The relevant electron scattering signatures are for Super-K (red), Borexino (orange), and LZ (green).  Note that the LZ electron recoil band is included for completeness but the neutrino detectors are found to have greater sensitivity.  
    }
    \label{fig:erecwindow}
\end{figure*}

We consider potential signals of atmospheric DM in both direct detection experiments and neutrino telescopes.  For DM direct detection, we utilize the analyses by LZ \cite{LZ:2018qzl} and PandaX-4T \cite{PandaX-4T:2021bab}. These experiments have the primary benefit of being sensitive to nuclear scattering.  Given that our production model is entirely baryonic, this allows for relevant constraints to be obtained in variants of our benchmark model which lack a DM-lepton coupling. The large boost provided by atmospheric DM production (see \cref{fig:diff}) also allows the use of neutrino observatories that have much higher recoil energy thresholds. We focus on dedicated electron scattering analyses by Borexino \cite{Borexino:2017uhp} and Super-Kamiokande (Super-K) \cite{Super-Kamiokande:2017dch} that can be applied directly to DM searches.
These detectors provide an advantage both in their overall size and in their much higher energy thresholds.  
For relativistic DM produced via bremsstrahlung, the detector kinetic energy thresholds (denoted $T_1$ and $T_2$)
of neutrino experiments in the $\mathcal{O}(\si{MeV}{-}\si{GeV})$ range are favourable compared to the $\mathcal{O}(10 \, \si{keV})$ thresholds for DM direct detection experiments.\\

We define the event spectrum as 
\begin{align}\label{eqn:eventdensity}
    \frac{d n}{dT_r}= \int_{E_\chi^\text{min}}^\infty dE_\chi \frac{d \Phi_\chi}{dE_\chi} \frac{d \sigma_{\chi r}}{d T_r},
\end{align}
where $d\sigma_{\chi r}/dT_r$ denotes the relativistic vector-mediated differential scattering cross section from \cite{deNiverville:2016rqh}, which gives the number of events per unit recoil kinetic energy $T_r$ where the label $r$ denotes either a recoiling nucleus or electron. Note that the bounds on $E_\chi$ are there for completeness, but the production model actually leads to a much narrower effective integration range.  The kinematic threshold imposed in \cref{eqn:gammacut} dictates that the DM flux becomes negligible far above the minimum scattering threshold, 
\begin{align}
    E_\chi^\text{min} = \frac{m_r}{m_r + m_\chi} T_r + m_\chi. 
\end{align}
The primary contribution to the flux in \cref{fig:diff} has $E_\chi$ between $1 \, \si{GeV}$ and $100 \, \si{GeV}$, thus a numerical integration cutoff is safely imposed at $E_\chi = 1000 \, \si{GeV}$.  The event spectrum is useful for visualizing the optimal recoil energy range for this process.  We observe in \cref{fig:erecwindow} 
that the event density is relatively flat as a function of kinetic recoil energy up to tens of MeV. Accordingly, large-volume neutrino experiments gain an advantage over conventional direct detection experiments due to their extended reach and larger exposures.

The event spectrum can be integrated to determine the total number of detected events, 
\begin{align}\label{eqn:events}
    N(m_\chi, \mathcal{R}, \epsilon) = \mathcal{E} \int_{T_1}^{T_2} dT_r \varepsilon(T_r) \frac{d n}{dT_r},
\end{align}
where $\mathcal{E}$ is the exposure and $\varepsilon(T_r)$ is the detector recoil efficiency with support $[T_1, T_2]$.  The shaded regions of \cref{fig:erecwindow} represent the recoil energy detection regions of the different detectors considered in this analysis.

\begin{figure*}[t]
    \centering
    \includegraphics{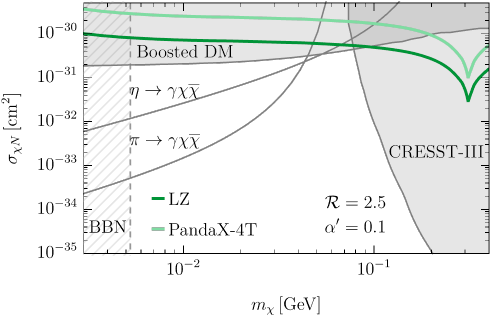}
    \includegraphics{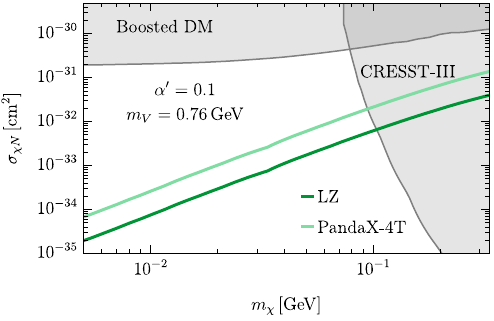}
    \caption{Constraints on the DM-nucleon cross section, from LZ and PandaX-4T sensitivity to vector-mediated DM produced  via atmospheric bremsstrahlung, compared to conventional direct detection limits from CRESST-III \cite{CRESST:2024cpr}. We also show the Xenon1T limits \cite{XENON:2018voc} applied to atmospheric production of DM via meson decay following \cite{Alvey:2019zaa} but incorporating appropriate partial decay rate kinematics for the on-shell decay $V \rightarrow \chi\bar{\chi}$. Similarly, the boosted DM contour is from Xenon1T limits on elastic up-scattering of DM by cosmic rays as calculated in \cite{Bringmann:2018cvk}. Finally, the hatched region is disfavoured for thermal DM by constraints from BBN.} 
    \label{fig:xsecN}
\end{figure*}

\begin{figure*}[t]
    \centering
    \includegraphics{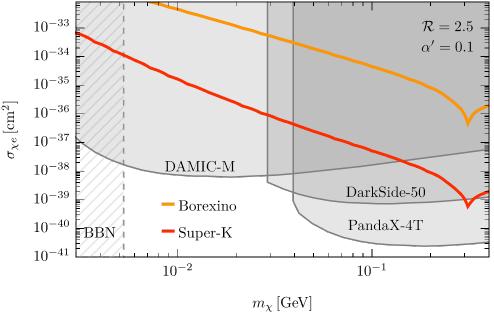}
    \includegraphics{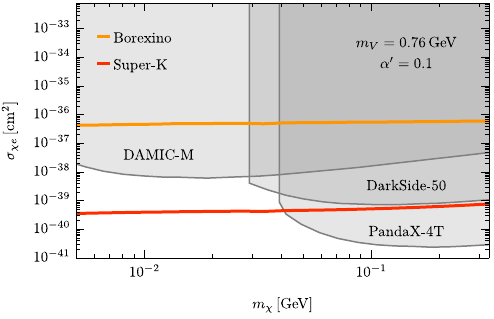}
    \caption{Constraints on the DM-electron cross section from Borexino and Super-K sensitivity to vector-mediated DM produced via atmospheric bremsstrahlung, compared to conventional direct detection limits on DM in the halo from PandaX-4T \cite{PandaX:2022xqx}, DarkSide \cite{DarkSide:2022knj}, and Damic-M \cite{DAMIC-M:2025luv}.}
    \label{fig:xsece}
\end{figure*}

The scattering rate of incident relativistic DM particles is  computed using the full differential cross section $d\sigma_{\chi r}/dT_r$ given in  \cite{deNiverville:2016rqh}. However, for ease of comparison with other direct detection constraints, the results are presented using the non-relativistic formalism, where we match 
\begin{align}\label{eqn:diffxsec}
    \frac{d \sigma_{\chi r}}{d T_r} = \sigma_{\chi r}\frac{m_r m_\chi}{4 \mu^2_{\chi r} T_\chi} \left\vert F(Q)\right\vert^2,
\end{align}
with $\mu_{ij}$ the reduced mass of particles $i$ and $j$ and $\sigma_{\chi r}$ the nonrelativistic reference cross section,
\begin{align}\label{eqn:refxsec}
    \sigma_{\chi r} = \frac{16 \pi \alpha \alpha' \epsilon^2 \mu_{\chi r}^2}{m_V^4}.
\end{align}
Constraints on $\epsilon$ from the full relativistic calculation are translated to $\sigma_{\chi r}$ using (\ref{eqn:refxsec}) when presenting our results in Figs.~\ref{fig:xsecN} and \ref{fig:xsece} below. Finally, the form factor $F(Q)$ is relevant for nuclear scattering with $Q$ the momentum transfer, and in that case we make use of a suitably normalized Helm form-factor for a nucleus $N$ with atomic mass $A$ \cite{Duda:2006uk}, 
\begin{align}
    \left\vert F(Q)\right\vert^2 = Z^2\left(\frac{\mu_{p\chi}}{\mu_{N \chi}}\right)^2\left(\frac{3 j_1(QR_1)}{Q R_1}\right)^2 e^{-Q^2 s^2},
\end{align}
where $R_1 \simeq (1.1\, {\rm fm}) A^{1/3}$ and $s \simeq 0.9\,{\rm fm}$.

We determine the 90\% confidence limit on the mixing parameter $\epsilon$ and thus the scattering cross section for each experiment following the procedure of \cite{Raj:2024guv}, that assumes Poisson event count statistics with a Gaussian distribution of background events,
\begin{align}
    \epsilon = \left(\frac{N_\text{cl}}{N(m_\chi, \mathcal{R}, 1)}\right)^{1/4}, \label{lim}
\end{align}
where 
\begin{align}
    N_\text{cl} = \frac{2}{\sqrt{\pi \sigma_\text{bkg}^2}}\left[1 + \erf\left(\frac{N_\text{bkg}}{\sqrt{2 \sigma_\text{bkg}^2}}\right)\right]^{-1}.
\end{align}
The scaling in \eqref{lim} is a consequence of the proportionalities $\frac{d \Phi_\chi}{d E_\chi}\propto\epsilon^2$ and $\frac{d \sigma_r}{d T_r}\propto\epsilon^2$.  

\begin{figure}[t]
    \centering
    \includegraphics{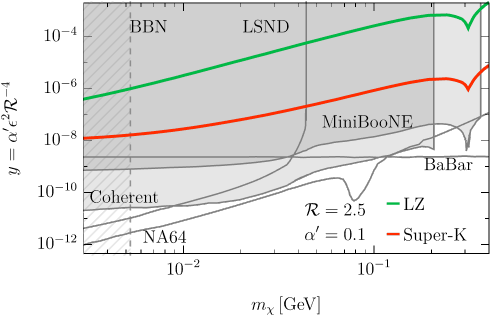}
    \caption{Constraints on $y = \alpha' \epsilon^2 / \mathcal{R}^4$ for our benchmark vector mediated DM model due to bremsstrahlung production in the atmosphere from LZ and Super-K.  We compare this sensitivity with competing fixed target and collider experiments. The contours for PandaX-4T and Borexino are similar to LZ and Super-K respectively, but are slightly weaker and are suppressed for clarity. Existing limits are filled for experiments with a necessary coupling to hadrons, while unfilled limits represent experiments that also rely on electron couplings.   We compare to limits from BaBar \cite{BaBar:2017tiz}, NA64 \cite{NA64:2023wbi}, Coherent \cite{COHERENT:2021pvd}, MiniBooNE \cite{MiniBooNEDM:2018cxm}, and LSND \cite{deNiverville:2011it, LSND:2001aii}. }
    \label{fig:epsilonexclusion}
\end{figure}

We present the final sensitivity results in terms of the reference cross-section defined in \cref{eqn:refxsec} when compared to the reach of direct detection experiments, and the parameter $y = \alpha' \epsilon^2 \mathcal{R}^{-4}$ for comparison with accelerator based experiments.  Note that although the reference cross-sections for nucleon and electron scattering are proportional to each other, we present the sensitivity contours separately here to highlight the comparative reach of experiments that rely on electron couplings, given that the proton bremsstrahlung production mode is purely hadronic. Note that simple generalizations of the model considered here (see e.g. \cite{Esseili:2025sbc}) do exhibit suppressed lepton couplings.

The most natural comparison is to the reach of existing direct detection experiments, exhibited in \cref{fig:xsecN} and \cref{fig:xsece}.  With a canonical correlated mass choice of $\mathcal{R} = m_V/m_\chi = 2.5$, the results are complimentary to existing limits from atmospheric DM production via meson production. For this parameter slice, the recently improved direct detection limit from CRESST-III excludes the advantageous resonant peak in the nuclear cross-section. The sensitivity contours for electron scattering in neutrino experiments are also excluded by a combination of direct detection electron scattering experiments for the canonical $\mathcal{R}=2.5$ slice of parameter space.

The DM production rate from proton bremsstrahlung is extremely sensitive to $m_V$. By choosing $m_V = 0.76\,\si{GeV}$, we optimize 
the output of the bremsstrahlung process due to vector mixing with the $\rho/\omega$ resonances (seen in \cref{fig:diff}).  The sensitivity of direct detection and neutrino experiments is substantially enhanced for this slice of parameter space, as exhibited in \cref{fig:xsecN} where the mediator mass becomes large enough that the meson decay channels are suppressed as the 3-body decay must proceed via an off-shell vector mediator.    

For completeness, we also show the more specific model-dependent sensitivity to the parameter $y$ for the $\mathcal{R}=2.5$ slice in \cref{fig:epsilonexclusion}. Sensitivity to atmospheric DM production is not competitive with current accelerator bounds. In particular, the BaBar limit on missing mass dominates in the $\rho/\omega$ resonance regime. However, it is notable that the leading accelerator limits and the atmospheric limit from Super-K all rely on the presence of electron couplings.

%%%%%%%%%%%%%%%%%%%%%%%%%%%%%%%%%%%%%%%%%%%%%%%%%%%%%%

%%%%%%%%%%%%%%%%%%%%%%%%%%%%%%%%%%%%%%%%%%%%%%%%%%%%%%
\section{Concluding Remarks}
\label{sec:disc}
%%%%%%%%%%%%%%%%%%%%%%%%%%%%%%%%%%%%%%%%%%%%%%%%%%%%%%

In this paper, we have revisited the sensitivity of DM direct detection experiments and neutrino telescopes to the atmospheric production of sub-GeV DM by cosmic ray interactions. We have extended the reach in mass by utilizing a recently developed model for the initial state radiation of dark vector mediators in proton-nucleon scattering. This has the advantage of form-factor resonant enhancements due to mixing with hadronic vector mesons. We indeed find enhanced sensitivity to benchmark models of sub-GeV DM interacting via a dark photon mediator with mass in the range of the broad $\rho/\omega$ resonance. For completeness, we illustrated the sensitivity for both the conventional $\mathcal{R} \equiv m_V/m_\chi = 2.5$ slice of parameter space, and the case with $m_V = 0.76 \, \si{GeV}$ closer to the peak of the $\rho/\omega$ resonance. While the ultimate sensivity to this specific benchmark vector mediated model is exceeded by terrestrial fixed target and collider experiments, the sensitivity over a specific range of parameter space exceeds that of direct detection limits on electron and nuclear scattering of DM in the halo. 

Moving forward, the continuous nature of atmospheric production naturally allows for the improvement in sensitivity with growing exposure. For dark photon mediated DM models, the possibility of electron scattering means that large volume neutrino telescopes have an advantage, as observed with the analysis of Super-K data in particular. Indeed, the sensitivity limits in this paper can naturally be extended for future large-volume neutrino detectors such as Hyper-K, DUNE and JUNO.

\section*{Acknowledgements}

We are grateful to F. Kling for helpful discussions. This research was supported in part by funding from NSERC, Canada, and the Arthur B. McDonald Canadian
Astroparticle Physics Research Institute. 

\bibliography{literature}

\end{document}